\documentclass[twocolumn,english,aps,amssymb,prl,showpacs]{revtex4}
\usepackage[T1]{fontenc}
\usepackage[latin9]{inputenc}
\usepackage{amsmath}
\usepackage{graphicx}
\usepackage{amssymb}
\usepackage{esint}

\makeatletter
\@ifundefined{textcolor}{}
{%
 \definecolor{BLACK}{gray}{0}
 \definecolor{WHITE}{gray}{1}
 \definecolor{RED}{rgb}{1,0,0}
 \definecolor{GREEN}{rgb}{0,1,0}
 \definecolor{BLUE}{rgb}{0,0,1}
 \definecolor{CYAN}{cmyk}{1,0,0,0}
 \definecolor{MAGENTA}{cmyk}{0,1,0,0}
 \definecolor{YELLOW}{cmyk}{0,0,1,0}
 }

\usepackage{epsfig}
\@ifundefined{showcaptionsetup}{}{%
\PassOptionsToPackage{caption=false}{subfig}}
\usepackage{subfig}
\makeatother

\usepackage{babel}

\begin{document}

\title{Chiral topological phases and fractional domain wall excitations
in one-dimensional chains and wires}

\author{Jukka I. V\"ayrynen}

\author{Teemu Ojanen}

\email[Correspondence to ]{teemuo@boojum.hut.fi}

\affiliation{Low Temperature Laboratory, Aalto University, P.~O.~Box 15100,
FI-00076 AALTO, Finland }

\date{\today}
\begin{abstract}
According to the general classification of topological insulators,
there exist one-dimensional chirally (sublattice) symmetric systems that can support any number of topological phases. We introduce
a zigzag fermion chain with spin-orbit coupling in magnetic field and
identify three distinct topological phases.
Zero-mode excitations, localized at the phase boundaries, are fractionalized: two of the
phase boundaries support $\pm e/2$ charge states while one of the
boundaries support $\pm e$ and neutral excitations. In addition, a finite chain exhibits $\pm e/2$ edge states for two of the three phases. We explain how the studied system generalizes the Peierls-distorted polyacetylene
model and discuss possible realizations in atomic chains and quantum
spin Hall wires.
\end{abstract}


\maketitle
\bigskip{}

\emph{Introduction}-- The discovery of quantum spin Hall effect in
two dimensional electron systems initiated an enormous wave of research
exploring topological properties of gapped band structures \cite{kane1,bernevig1}.
This has lead to various recent discoveries of novel states of matter,
commonly known as topological insulators, exhibiting striking properties
such as exotic electromagnetic responses and gapless surface states \cite{kane2}.
The quest for novel topological insulators, guided by the comprehensive classification presented in Ref.~\cite{schnyder}, is one of the
most active topics in condensed matter physics at the moment.


In one spatial dimension (1D) the recent attention has concentrated
on the topological superconducting phases realized in spin-orbit coupled
semiconductor nanowires in proximity of a supeconductor \cite{lutchyn,oreg}.
It was proposed that networks of wires could serve as a
platform for topological quantum computation \cite{alicea}. However,
the study of topological phases in 1D systems can be traced back over
30 years to the discovery of soliton excitations in polyacetylene \cite{su1}.
Indeed, the Su-Schrieffer-Heeger model is an example of an
insulator exhibiting a chiral symmetry \cite{gurarie}.
The zero-energy states located between differently dimerized domains
support neutral spin-$1/2$ excitations and spinless $\pm e$ excitations.

In this work we introduce a zigzag fermion chain model with a spin-orbit
coupling in a magnetic field. This system belongs to the BDI class
in the classification scheme and has three distinct topological phases depending on the orientation
and magnitude of the magnetic field. The domain wall (DW) excitations
localized at the phase boundaries are fractionalized: two of the DWs
support a single zero mode with possible charge states $\pm e/2$
while one DW supports two zero modes leading to $\pm e$ and neutral excitations. Furthermore, a finite chain exhibits $\pm e/2$ charged edge states for two of the three phases. The most striking differences to the polyacetylene model include existence of three distinct phases (instead of two) and possibility to realize $\pm e/2$ excitations that are absent in polyacetylene due to spin degeneracy.
Zigzag chains are realized in a number of systems which indicates
that the studied model could be relevant, for example, in
monoatomic gold chains in magnetic fields. Additionally, a continuum
limit of the considered model describes narrow wires of quantum spin
Hall insulators. Interestingly, in this
case the $\pm e/2$ excitations can be thought of as a bound pairs
of $\pm e/4$ charges on opposite edges of the wire.

\begin{figure*}[t]
\raggedright{}\subfloat[\label{fig:chainandqsh}]{\includegraphics[width=0.4\columnwidth, clip=true]{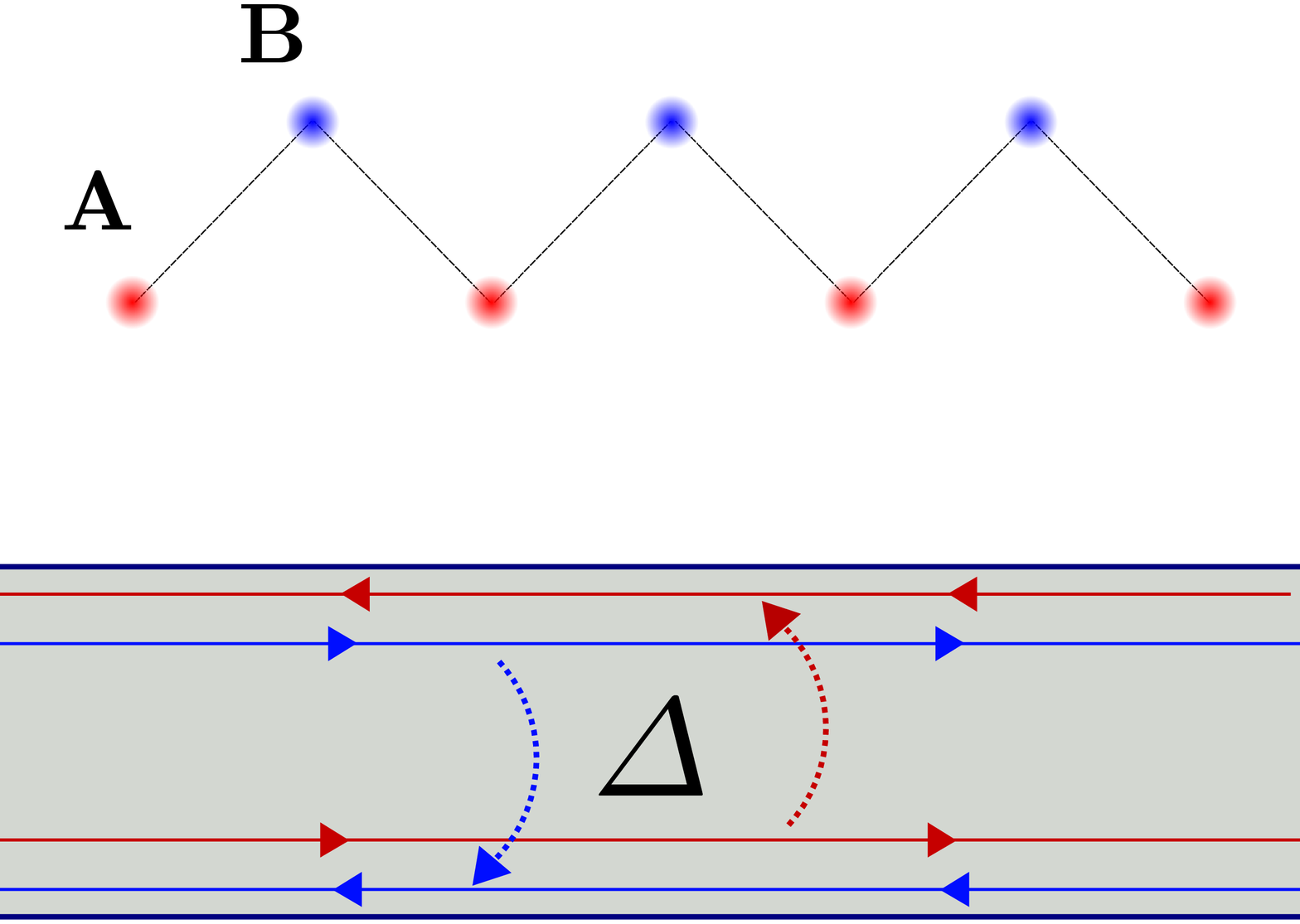}}\hspace*{\fill}\subfloat[\label{fig:delta0phasediag}]{\includegraphics[height=0.35\columnwidth]{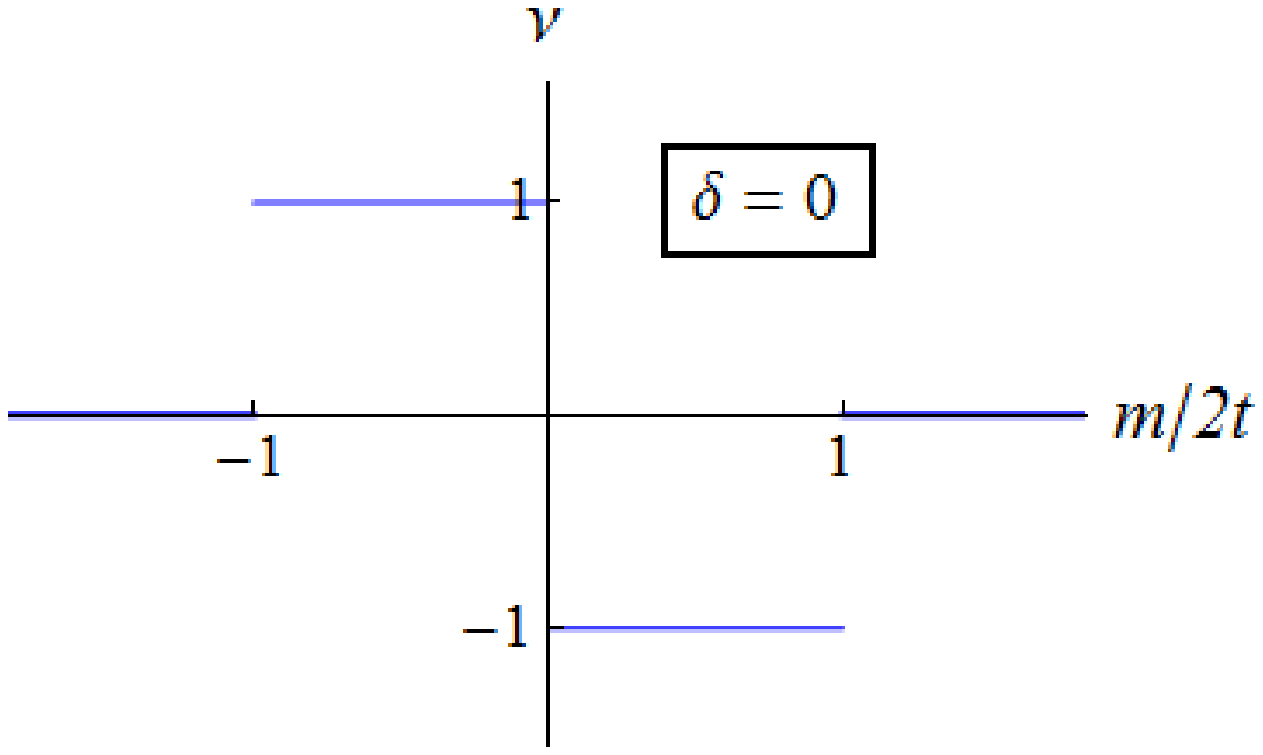}}\hspace*{\fill}\subfloat[\label{fig:deltafinitephasediag}]{\includegraphics[height=0.35\columnwidth]{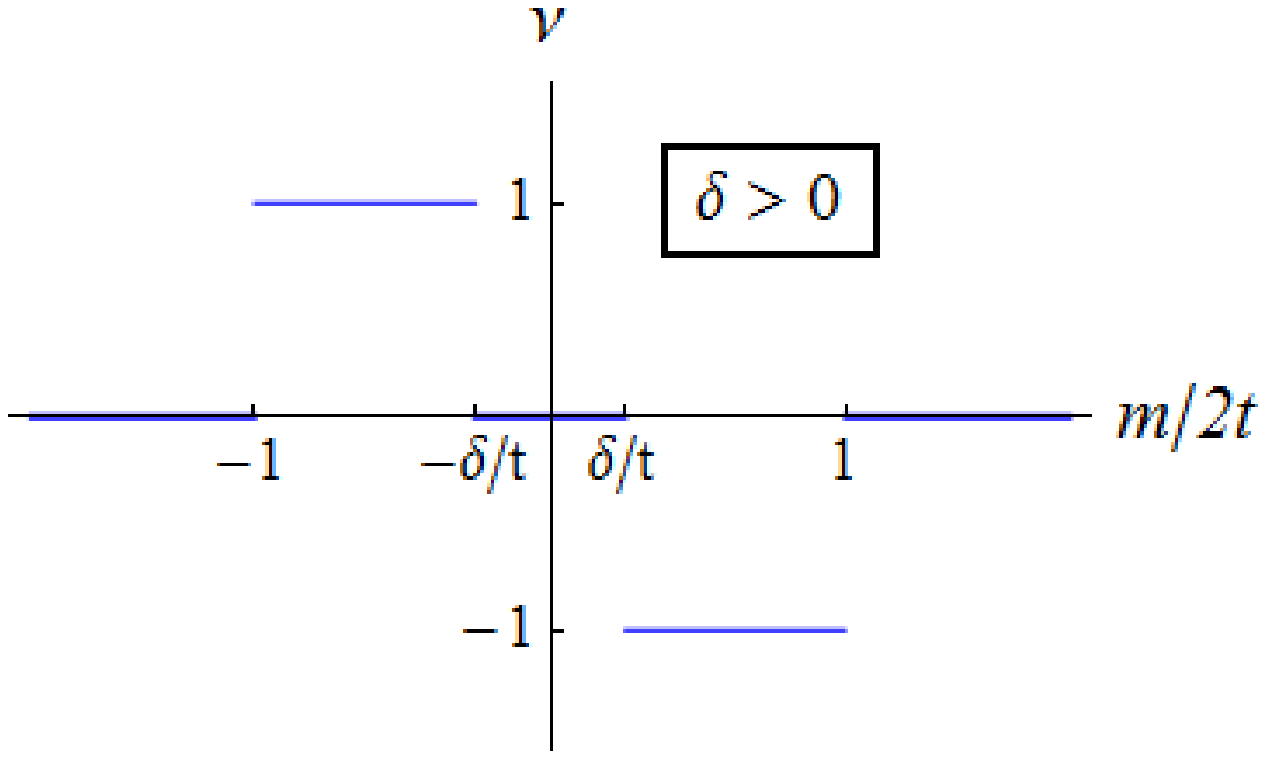}}\hspace*{\fill}\subfloat[\label{fig:qshphasediag}]{\includegraphics[height=0.35\columnwidth]{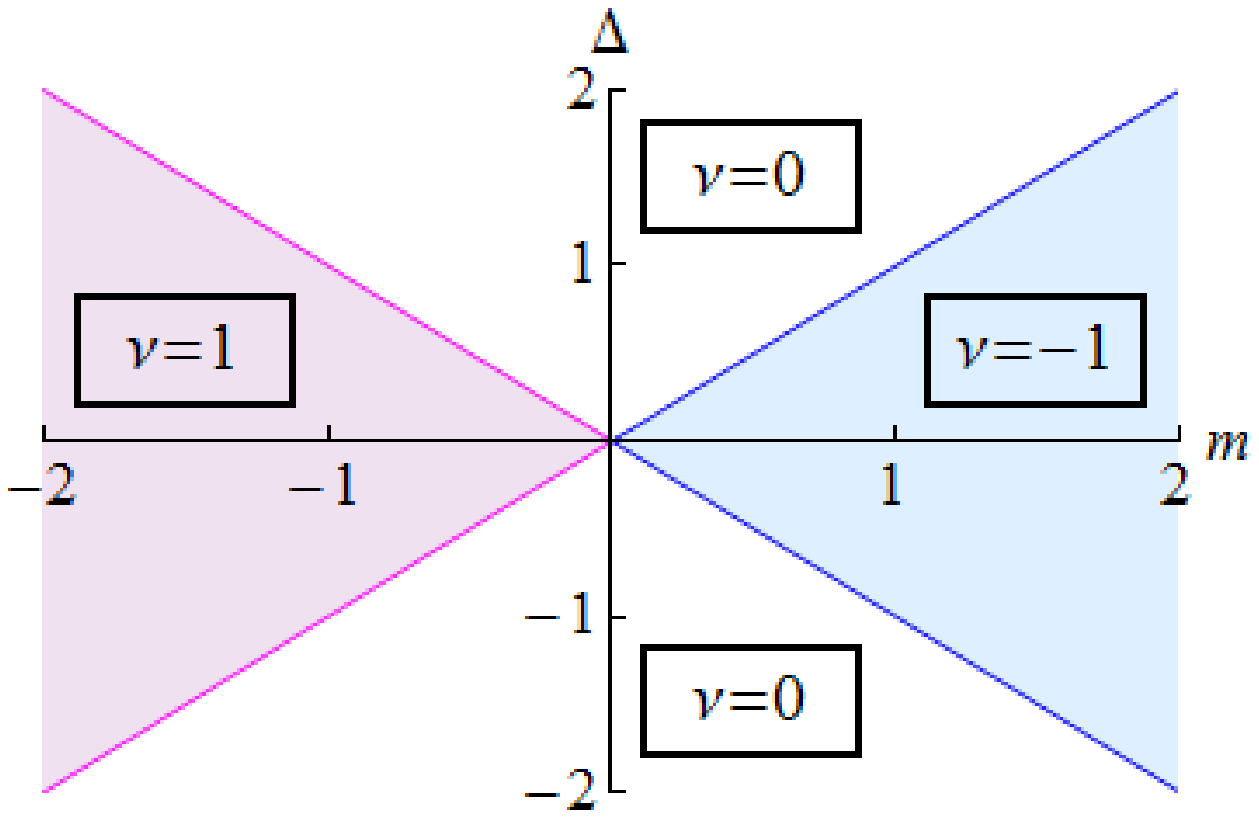}}
\caption{\label{fig:allphasediags}(a) Studied 1D models, a zigzag fermion chain with a spin-orbit hopping and narrow quantum spin Hall wire with a finite inter-edge coupling. Topological properties of these systems are closely related. (b--d) The phase diagrams of Hamiltonians (\ref{bloch}) and (\ref{QSH}) as a function of model parameters. (b) For vanishing hopping modulation $\delta=0$ there is a ($1/-1$) interface at $m=0$. (c) For a finite $\delta$, a trivial phase emerges for $|m|<|\delta|$. (d) The Hamiltonian (\ref{QSH}) shows three different phases for finite inter-edge coupling $\Delta$. The normalization of (\ref{wind}) is multiplied by 2 since continuum models produce half-integer values.}
\end{figure*}

\emph{1D models and phase diagrams}-- In strictly 1D chains the
Peierls instability commonly results in geometrical deformations doubling
the unit cell and creating a sublattice structure. If atoms are
not restricted to a line, chains can relax to planar zigzag or more
complicated structures. Motivated by this, we consider a zigzag fermion
chain shown in Fig.~\ref{fig:chainandqsh}. The Hamiltonian of the
system is \begin{align}
H & =t\sum_{\langle i,j\rangle}c_{i}^{\dagger}c_{j}+\delta\sum_{<i,j>}(-1)^{i}c_{i}^{\dagger}c_{j}+\nonumber \\
& i\frac{\lambda}{|\boldsymbol{e}_{ij}|}\sum_{\langle\langle
i,j\rangle\rangle}c_{i}^{\dagger}\mathbf{S}\cdot\boldsymbol{e}_{ij}c_{j}+\sum_{i}c_{i}^{\dagger}\,\boldsymbol{m}\cdot\mathbf{S}\, c_{i}\label{eq:Hamiltonian}\end{align}
where the first term describes the nearest-neighbor hopping and the second
term encodes the hopping modulation due to dimerization.
The third term arises from the next-nearest neighbor spin-orbit hopping
and the last term corresponds to the Zeeman splitting $\boldsymbol{m}=(m_{x},m_{y})$ due to the external
magnetic field. The spin-orbit term
depends on the orientation of the two bonds connecting the next-nearest
neighbors $\boldsymbol{e}_{ij}=\boldsymbol{d}_{i}\times\boldsymbol{d}_{j}$
and vanishes for strictly linear chains \cite{kane1}. The spin-orbit coupling and magnetization are crucial for the studied topological properties of the system
while the hopping modulation is included for the sake of generality. The model (\ref{eq:Hamiltonian}),
with an in-plane magnetic field leads to the Bloch Hamiltonian
\begin{align}
H(k)=2t\,\mathrm{cos}(ka)\sigma_{x}+2\delta\,\mathrm{sin}(ka)\sigma_{y}+\nonumber \\
+2\lambda\,\mathrm{sin}(2ka)\sigma_{z}S_{z}+mS_{x}\label{bloch}\end{align}
where Pauli matrices $\sigma_{i}$ and $S_{i}$ operate in the sublattice
and spin space, respectively. The Hamiltonian ($\ref{bloch}$) possesses
a chiral symmetry: it anticommutes with $C=\sigma_{z}S_{y}$. This
property has important consequences on the energy spectrum and topological
properties of the system \cite{schnyder}. If $H$ has an eigenstate
$|E_{i}^{+}\rangle$ with energy $E_{i}$, there exists another eigenstate
$C|E_{i}^{+}\rangle$ belonging to the energy $-E_{i}$, so the spectrum
is symmetric about $E=0$. Zero modes, if
existing, can be chosen to coincide with the eigenstates of the generator
of the chiral symmetry, $C|\psi^{\pm}\rangle=\pm|\psi^{\pm}\rangle$.
In odd spatial dimensions systems with a chiral symmetry may posses
topologically nontrivial phases characterized by a $Z$-valued topological
index  introduced in Ref.~\cite{schnyder}. Here we employ alternative but equivalent
formulation \cite{volovik}.
In 1D the invariant takes the form
\begin{align}
\nu[H]=\frac{1}{8\pi i}\int dk\,\mathrm{Tr}\left[CH^{-1}\partial_{k}H\right],\label{wind}\end{align}
where the integration extends over the Brillouin zone for Bloch Hamiltonians
and over the real line for continuum models.

In Fig. \ref{fig:allphasediags} we plot the invariant (\ref{wind})
for Hamiltonian (\ref{bloch}) as a function of the magnetic field,
revealing the phase diagram. The system has three distinct phases
$\nu=\pm1$ and $\nu=0$, separated by quantum phase transitions.
In these interfaces where the invariant changes its value, there exist
zero energy states protected by the chiral symmetry. The phase of
the system depends on $m$ and can thus be changed by tuning the magnitude
of the external field, i.e., by introducing magnetic DWs into the
system. At large field values $(|m|>2t)$ the magnetization dominates
and $\nu=0$. We will show below that this phase can be identified with vacuum.


At vanishing spin-orbit coupling $\lambda=0$ and magnetization
$\boldsymbol{m}=0$, model (\ref{bloch}) describes a Peierls-distorted polyacetylene chain \cite{su1}.
The resulting Hamiltonian anticommutes with $C'=\sigma_{z}$ which serves as
the generator of the chiral symmetry. Evaluation of the invariant
(\ref{wind}) reveals two distinct topological phases with $\nu=\mathrm{sign}\,\delta$.
These phases correspond to the well-known degenerate dimer configurations
of the chain. There is, however, important differences between the
polyacetylene model and the zigzag chain. In polyacetylene there is no
$\nu=0$ phase which in our model is identified with vacuum. As a consequence, a finite polyacetylene chain does not exhibit robust edge states unlike $\nu=\pm1$ zigzag chains. In addition, due to
the spin degeneracy in polyacetylene, there are always two zero modes at phase
boundaries. As we emphasize below, due
to this fact the zigzag chain exhibits fractional domain wall charges
$\pm e/2$ whereas polyacetylene has $\pm e$ and neutral excitations.

There is an interesting connection between the model (\ref{bloch})
and quantum spin Hall (QSH) wires (Fig. \ref{fig:chainandqsh}). A
QSH bar supports helical edge states where the velocity and spin of
the edge carriers are completely correlated \cite{kane2}. In a narrow enough wire
the edge modes localized on opposite edges begin to overlap, enabling
interedge hybridization which gaps the edge spectrum. The Hamiltonian
describing the systems is
\begin{align}
H(k)=v_{F}k\,\sigma_{z}S_{z}+\Delta\sigma_{x}+mS_{x},\label{QSH}\end{align}
where the first term describes the unperturbed edge states, the second
term comes from the inter-edge tunneling and the last term corresponds to
magnetization. The Pauli matrices $\sigma$ now operate
in the subspace of the two edges of the wire. Setting $\delta=0$
and expanding the band Hamiltonian (\ref{bloch}) around $k=0$ or
$k=\pi$, and identifying $2t\to\Delta$ and $4\lambda a\to v_{F}$,
one recovers (\ref{QSH}). The close connection between (\ref{bloch})
and (\ref{QSH}) suggests that the topological properties of these
systems bear similarities. The phase diagram of (\ref{QSH}), shown
in Fig. \ref{fig:qshphasediag}, reveals three distinct phases as
a function of the relevant control parameters $\Delta$ and $m$.
The zero-energy states at DWs are also very similar to those of the
zigzag chain as we explain below.

\emph{Zero modes and fractional excitations}-- Now we analyze
zero modes by numerical diagonalization of the lattice model and analytical
methods and identify associated quantum numbers. As mentioned above the spectrum is symmetric with respect to $E=0$.
This property and the number of the zero modes at DWs are sufficient to
determine the charge states of the zero mode excitations \cite{su1,jackiw}. According to the general argument, one-half of an
electron state per zero mode is missing from the valence band in the
vicinity of a DW. If a DW supports a single zero mode, then the possible
charge states at half filling are $\pm e/2$ depending on the population of the zero mode. If there exist two zero modes, possible charge
states are $\pm e$ and $0$. The total
energy of the system is independent of the population of the zero
modes. In finite systems where the DWs come in pairs zero modes are
split to linear combinations of $|\psi^{\pm}\rangle$ with finite energies around $E=0$,
approaching zero as the separation of the DWs increases.

\begin{figure}
\subfloat[\label{fig:wavefunctions}]{\includegraphics[width=0.67\columnwidth]{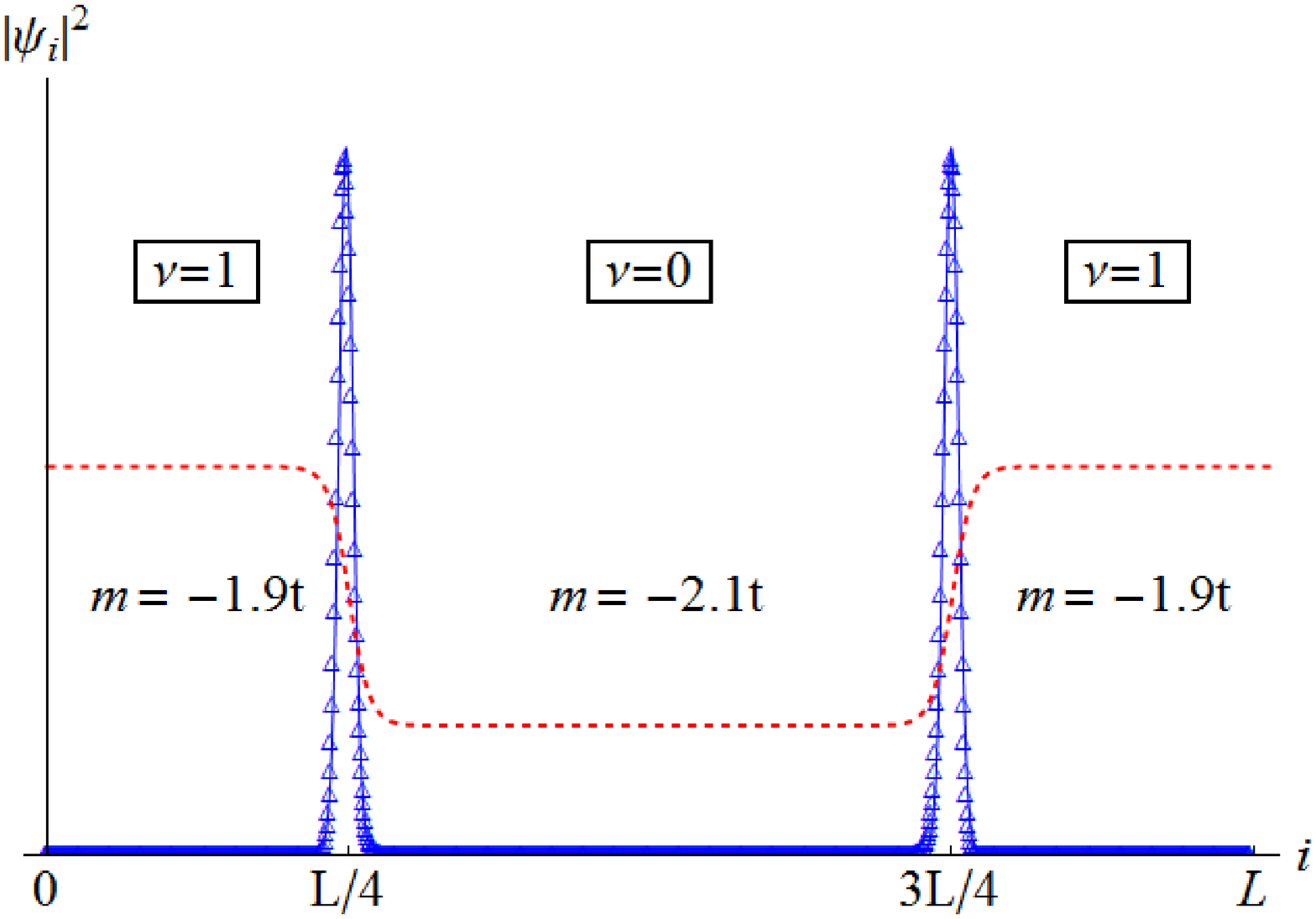}}\hspace*{\fill}\subfloat[\label{fig:QSHeper4}]{\includegraphics[width=0.3\columnwidth, clip=true]{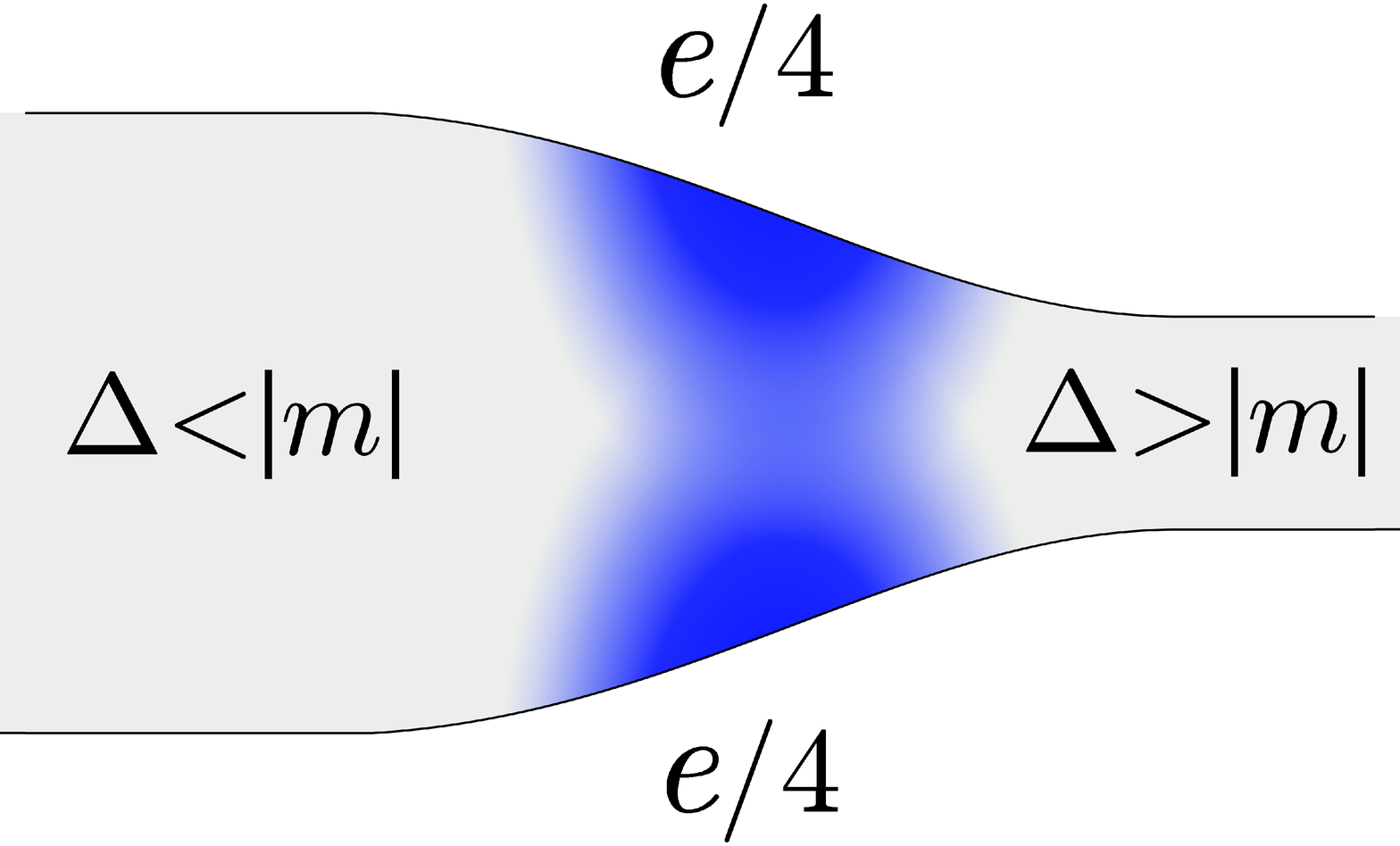}}
\caption{(a) Numerical solution of zero mode wavefunctions (blue triangles) for a ring with $L=1200$ sites, $\lambda/t = 1/8$ and an analytical approximation based on continuum solutions (\ref{zero1}). The red dashed line represents the magnetization profile. (b) In QSH wires a 1/0 DW supports a fractional charge state $\pm e/2$ distributed evenly between $\pm e/4$ charges.}
\end{figure}

At the DWs where the phase $\nu=1$ changes to $\nu=0$ (referred
as the 1/0 DW) we solve the model (\ref{eq:Hamiltonian}) in a finite
ring with spatially varying magnetization $m$. For simplicity we assume that $\delta=0$ in this section, so there are two phase
boundaries located at places where $m$ crosses $2t$. We choose a
magnetization profile so that the DWs occur at the opposite sides
of the ring. As illustrated in Fig.~\ref{fig:wavefunctions}, we discover two mid-gap energy
states, both located at the DWs. When the ring size is increased the
energies move closer to zero and eventually become degenerate. This
result can be complemented by an analytical calculation based on the continuum approximation of Eq.~(\ref{bloch}).
Linearizing (\ref{bloch}) at Brillouin zone point $k=0$ (or $\pi$) where
the gap closes, we solve the Schr\"odinger equation for zero energy
states. We find one normalizable solution $\psi_{1/0}(x)$ ($\psi_{0/1}(x)$)
located at a DW  (anti-DW),
\begin{align}
&\psi_{1/0}(x)=A\, e^{\int_{x_{0}}^{x}dx'\frac{2t+m(x')}{4\lambda\, a}}|a^{+}\rangle \nonumber\\
&\psi_{0/1}(x)=A\, e^{-\int_{x_{0}}^{x}dx'\frac{2t+m(x')}{4\lambda\, a}}|a^{-}\rangle \label{zero1}
\end{align}
where $|a^{\pm}\rangle=\left(|\uparrow\rangle|y_{\pm}\rangle+|\downarrow\rangle|y_{\mp}\rangle\right)$
and $|\uparrow,\downarrow\rangle$, $|y_{\pm}\rangle$ denote the
eigenstates of $\sigma_{z}$ and $S_{y}$. The DW profile is determined
by $m(x)$. For $\psi_{1/0}$ ($\psi_{0/1}$) $m(x)$ is an
arbitrary decreasing (increasing) non-vanishing function satisfying
$m(x_{0})=-2t$ and approaching constant value away from $x_{0}$.
The normalization constant $A$ is determined by the detailed form
of $m(x)$. According to the general argument, there is a charge
aggregate of $\pm e/2$ localized at the DW depending on whether the
zero mode is populated ($-e/2$) or not ($e/2$) \cite{su1}. Since the total
number of electrons must be an integer, the DWs must come in \textquotedbl{}soliton-antisoliton\textquotedbl{}
pairs which is clear in the ring geometry. In a finite linear
chain in the phase $\nu=\pm1$ terminated by hard-wall boundary conditions
we discover one zero mode at each end of the chain. Thus the terminated end
behaves like the $\pm1$/$0$ interface and supports $\pm e/2$ excitations. A finite
chain in the phase $\nu=0$ does not
exhibit mid-gap states, indicating that the $\nu=0$ phase can indeed be
identified with the vacuum. The corresponding DW states for model (\ref{QSH}) can be obtained from Eq.~(\ref{zero1}) by substituting $2t\to\Delta$ and $4\lambda a\to v_{F}$. The $-1/0$ DW behaves similarly to the 1/0
DW and does not require a separate treatment.

In contrast to $\pm1/0$ DWs, a $-1/1$ DW supports two zero modes.
The numerical diagonalization reveals four mid-gap states (two per DW) approaching zero energy as the
ring size is increased. The wavefunctions of these states are localized
at the DWs. Analytical approximation for the wavefunctions are obtained by linearizing (\ref{bloch}) around $k=\pi/2$ and solving
DW $\psi_{1/-1}(x)$ and anti-DW  $\psi_{-1/1}(x)$ states
\begin{align}
\psi_{-1/1}(x)= & A_{\pm}\, e^{\int_{x_{0}}^{x}dx'm(x')\frac{-2ti\pm4\lambda}{a\,((2t)^{2}+(4\lambda)^{2})}}|b^{\pm}\rangle\nonumber \\
\psi_{1/-1}(x)= & B_{\pm}\, e^{\int_{x_{0}}^{x}dx'm(x')\frac{2ti\pm4\lambda}{a\,((2t)^{2}+(4\lambda)^{2})}}|c^{\pm}\rangle, \label{zero2}
\end{align}
where $|b^{\pm}\rangle=\left(|\uparrow\rangle|y_{+}\rangle\mp |\downarrow\rangle|y_{-}\rangle\right)$, $|c{\pm}\rangle=\left(|\uparrow\rangle|y_{-}\rangle\pm|\downarrow\rangle|y_{+}\rangle\right)$ .
For $-1/1$ DW ($1/-1$) $m(x)$ is increasing (decreasing) function
satisfying $m(x_{0})=0$ and approaching constant value away from
$x_{0}$. Since there are two zero modes per DW (anti-DW), the possible
charge states are, as in polyacetylene, $\pm e$ and 0 depending whether
both of the states are empty ($e$), one of them is excited (0) or
both are excited ($-e$). In the case of an antisymmetric DM $m(x-x_0)=-m(x_0-x)$ the valence band is spin paired so $\pm e$ charge states are actually spinless as in polyacetylene \cite{su1}. The argument for DW charge states is supported by calculations based on a method introduced in Ref.
\cite{vayrynen} recently. After an appropriate regularization, the DW charge (modulo multiples of $e$) follows from the expression
\begin{equation}
q=\frac{e}{4\pi^{2}}\int_{-\infty}^{\infty}dx\int_{-\infty}^{\infty}d\omega\int dk\mathcal{G}\partial_{k}\mathcal{G}^{-1}\mathcal{G}\partial_{\omega}\mathcal{G}^{-1}\mathcal{G}\partial_{x}\mathcal{G}^{-1}\nonumber\end{equation}
where $\mathcal{G}=(i\omega-H(k,x))^{-1}$ is the Wigner
transformed Green function \cite{gurarie,vayrynen}. To extract
DW charge, we will use the continuum approximation of $H$ in the vicinity of the DW.
Evaluation of the DW charge formula confirms that possible charges states for a $\pm1/0$ DW ($-1/1$) are half-integral (integral) multiples of $e$.

The possible charge states of $1/0$ and $1/-1$ DWs coincide with spinless and spinfull polyacetylene DWs.
However, the corresponding wavefunctions are quite different from those of polyacetylene zero modes, where DW and anti-DW wavefunctions
are nonzero only on separate sublattices \cite{su1}. In our model the sublattice and spin
structure follows directly from the property $C|\psi^{\pm}\rangle=\pm|\psi^{\pm}\rangle$,
where the different signs correspond to DW and anti-DW zero modes. Our numerical calculations indicate that mid-gap states are insensitive
to even-odd variation in the number of lattice sites and robust against
weak disorder in system parameters.  

\emph{Physical realizations}-- Zigzag fermion chains are known to
exist in various physical contexts. Structurally the most closely related
examples are monoatomic Au, Pt, and Ir chains \cite{smit,seivane}. For example,
Au spontaneously forms zigzag chains with one s-type conduction electron
per site. Since spin-orbit effects are pronounced in heavy elements,
these metal chains could provide a realization of the studied phenomenon.
The exact shape of these structures, magnitude of the spin-orbit
hopping $\lambda$ and the $g$-factor determining $m$ depend
on the details of the system and are not readily available. Thus it is difficult to give a realistic
estimate for the energy gap setting the relevant temperature scale. However, if our model captures the essential physics for these systems, a finite chain exhibits $\pm e/2$ edge states for any finite $\lambda$ and $|m|<2t$ at low temperatures.

More concrete estimates can be given for the QSH nanowire model (\ref{QSH})
which has similar topological properties. In a thin wire the edge states on opposite edges have a weak
overlap, inducing a gap $\Delta$ in the edge spectrum (\ref{QSH}).
This gap depends sensitively on the width of the wire \cite{zhou} and can be engineered by modulating it.
Magnetic field can be employed to tune the magnetization
$m$ \cite{qi} so by varying $m$ and $\Delta$ it is possible to drive
the system to different phases illustrated in Fig. \ref{fig:qshphasediag}.
In Ref.~\cite{zhou} it was estimated that for wires of width 200 nm the inter-edge
coupling is $\Delta=5.22$ K, while in Ref.~\cite{qi} it was concluded
that in-plane magnetization of $m=3.5$ K can realized by 
magnetic fields of the order 1T. These values suggest that experimental investigation
of the phase diagram and the zero mode excitations of QSH wires is
possible within existing methods at sub-Kelvin temperatures and high
magnetic fields. As discussed in Ref.~\cite{qi}, the single-flavor
Dirac spectrum realized in the edges of QSH systems have $\pm e/2$
excitations located at magnetic DWs. In a QSH wire the $\pm1/0$ phase boundary realizes another
exotic possibility, bound pairs of $\pm e/4$ charges on opposite
sides of the wire, as illustrated in Fig. \ref{fig:QSHeper4}. The zero
modes $(\ref{zero2})$ are localized equally in both sublattices indicating that equal charges reside on opposite edges of the wire. Since
the total charge on a DW is $\pm e/2$, opposite edges must have equal
charges $\pm e/4$. The $\pm1/0$ DW is taking place when magnetization
and tunneling are equal, determining the critical width of the wire. In the narrow
(wide) part of the wire tunneling can be made stronger (weaker) than
the magnetization, resulting in the $0/1$ DW when the wire width
exceeds the critical width. Unlike $e/2$ charges studied in \cite{qi},
$e/4$ charges rely on inter-edge coupling and are inherently bound together. The maximum distance of $e/4$ charges is limited
by the critical width of the wire where $m$ exceeds $\Delta$. Still,
for wires of width 200 nm or wider the charges have a very small overlap
and can be thought of as separate entities.%

\emph{Conclusion}-- We introduced a realistic 1D fermion chain exhibiting
three topological phases protected by chiral symmetry. Two of the
phase boundaries support $\pm e/2$ zero modes while one of the phase
boundaries have $\pm e$ and neutral excitations. We showed that similar
topological properties are shared by quantum spin Hall wires in magnetic
fields. We solved the zero energy states using combination of analytical
and numerical methods and discussed how predicted properties could
be realized in atomic chains and quantum spin Hall wires.

The authors would like thank Grigory Volovik, Hans Hansson and Jens Bardarson for valuable discussions.
This work was supported by Academy of Finland (T.O) and ERC Grant
No. 240362-Heattronics (J.I.V).

\end{document}